\documentclass[aps,prb,twocolumn,superscriptaddress,showpacs]{revtex4}
\usepackage{graphics}
\usepackage{graphicx}
\usepackage[english]{babel}
\usepackage{color}

\begin{document}

\title{Persistence of the first-order  transition lines in
mesoscopic Bi$_{2}$Sr$_{2}$CaCu$_{2}$O$_{8}$ vortex matter with
less than hundred vortices}

\author{M. Konczykowski}
\affiliation{Laboratoire des Solides Irradi\'{e}s, Ecole
Polytechnique, CNRS URA-1380, 91128 Palaiseau, France}

\author{Y. Fasano$^*$}
\affiliation{Laboratorio de Bajas Temperaturas, Centro At\'{o}mico
Bariloche, 8400 Bariloche, Argentina}

\author{M.I. Dolz}
\affiliation{Departamento de F\'{i}sica, Universidad de San Luis,
5700 San Luis, Argentina}

\author{H. Pastoriza}
\affiliation{Laboratorio de Bajas Temperaturas, Centro At\'{o}mico
Bariloche, 8400 Bariloche, Argentina}

\author{V. Mosser}
\affiliation{Itron SAS, F-92448 Issy-les-Moulineaux, France}

\author{M. Li}
\affiliation{Kamerlingh Onnes Laboratorium, Rijksuniversiteit
Leiden, 2300 RA Leiden, The Netherlands}

\date{\today}

\begin{abstract}
The persistence of the first-order  transition line in the phase
diagram of \textit{mesoscopic} Bi$_{2}$Sr$_{2}$CaCu$_{2}$O$_{8}$
vortex matter is detected down to a system size of less than
hundred vortices. Precise and highly-sensitive to bulk currents
\textit{AC} magnetization techniques proved to be mandatory in
order to obtain this information. The location of the vortex
matter first-order transition lines are
 not altered by decreasing the sample size down to  20\,$\mu$m.
 Nevertheless, the onset of irreversible magnetization is
 affected by increasing the sample surface-to-volume ratio producing
 a noticeable enlargement of the irreversible vortex region above
 the second-peak transition.
\end{abstract}

\pacs{74.25.Uv,74.25.Ha,74.25.Dw} \keywords{}

\maketitle

\section{Introduction}

The vortex matter phase diagram in macroscopic samples of clean
layered high-temperature superconductors presents a first-order
transition (FOT) line \cite{Pastoriza94a,Zeldov95a} between a
solid phase at low fields and temperatures and a liquid
\cite{Nelson1988} or decoupled gas \cite{Glazman1991} of pancake
vortices with reduced shear viscosity \cite{Pastoriza1995}. At low
temperatures the solid phase presents irreversible magnetic
behavior ascribed to bulk pinning and surface barriers, each of
them dominating at different temperature  and measuring-time
ranges \cite{Chikumoto1991,Chikumoto1992,Zeldov1995b}. Direct
imaging of the low-field phase in pristine samples of the
extremely anisotropic Bi$_{2}$Sr$_{2}$CaCu$_{2}$O$_{8}$ compound
reveals a vortex solid with quasi long-range positional order
\cite{Fasano2005,Fasano2008}. Heavy-ion-irradiated
Bi$_{2}$Sr$_{2}$CaCu$_{2}$O$_{8}$ samples also present a FOT at
low fields though in this case the vortex solid is polycrystalline
\cite{Banerjee2003,Menghini2003}. These findings indicate that the
positional order of the solid phase is therefore not relevant for
the order of the transition \cite{Menghini2003}.
Josephson-plasma-resonance measurements reveal the FOT corresponds
to a single-vortex decoupling process between stacks of pancake
vortices in adjacent CuO planes \cite{Colson2003}. Therefore,
there is evidence that the first-order transition is actually a
single-vortex transition that depends, at best, on the density of
the surrounding vortex matter. This can be further tested by
decreasing the sample size down to a few micrometers such that, at
low fields, the system size is reduced to only a few vortices.

The FOT of vortex matter in Bi$_{2}$Sr$_{2}$CaCu$_{2}$O$_{8}$ can
be detected at high temperatures, $0.66\,T_{\rm c}\lesssim T
<T_{\rm c}$, as a jump in the local flux density, or, in some
cases, in the reversible magnetization \cite{Zeldov95a}, or,
alternatively, by a frequency-independent paramagnetic peak in the
in-phase component of the first-harmonic of the AC screening
signal \cite{Morozov1996}. Both features develop at the same
field, identified with the first-order transition field $H_{\rm
FOT}$. The irreversible magnetic response of the system in this
high-temperature regime is dominated by a surface barrier.
Depending on the sample geometry, this barrier can be of the
geometrical or  Bean-Livingston type
\cite{Chikumoto1992,Morozov1996b}, while the bulk of the sample
remains magnetically reversible \cite{Zeldov1995b}. In the
intermediate temperature region $0.39\,T_{\rm c}\lesssim T
\lesssim 0.66\,T_{\rm c}$, bulk pinning increases its relevance on
cooling and the paramagnetic peak is masked  by bulk shielding
currents, entailing a sudden decrease of the AC signal
\cite{Morozov1996}. Finally, in the low-temperature regime $T
\lesssim 0.39\,T_{\rm c}=35\,$K, bulk pinning plays a dominant
role and the so-called second-peak effect
\cite{Avraham2001,Chikumoto1991} or order-disorder transition
\cite{Khaykovich97a,Vinokur1998,Cubbit1993} is detected. This
transition is manifested as a local increase of the width of DC
hysteresis loops \cite{Chikumoto1991}, and as a minimum in AC
magnetization loops with the field-location of the feature,
$H_{\rm SP}$, being frequency-independent
\cite{Sochnikov2008,vanderBeek1996}.

The persistence of these transition lines when decreasing the
sample size down to few vortices is still open to discussion.
Whether the thermodynamic $H_{\rm FOT}$ transition remains in
mesoscopic vortex matter consisting of only a few vortices has, to
our knowledge, not yet been investigated. Regarding the $H_{\rm
SP}$ transition, two works of the same authors have reported on
its disappearance  on decreasing the sample system size down to
30\,$\mu$m \cite{Wang2000,Wang2001}. The authors attributed this
phenomenology to the sample size becoming smaller than the
temperature-dependent Larkin-Ovchinikov correlation length
\cite{Larkin1979} down which the vortex structure is insensitive
to the  pinning potential. This interpretation is
 contradictory with their own  data since they invoke bulk
current arguments for a system that clearly presents a
surface-barrier-dominated physics as deduced from the two-quadrant
locus of their DC magnetization loops \cite{Wang2000,Wang2001}.
  Other work reports on a sample-size-dependent
$H_{\rm SP}$, even when not extremely varying the millimeter-range
sample size, attributed to a distribution of metastable disordered
vortex states with different lifetimes \cite{Kalisky2005}.

In this work we report on the phase-location of the $H_{\rm FOT}$,
$H_{\rm SP}$, and irreversibility, $H_{\rm irr}$, lines for
\textit{macroscopic} and \textit{micron-sized}
Bi$_{2}$Sr$_{2}$CaCu$_{2}$O$_{8}$ single crystals, with the field
applied parallel to the sample c-axis. We show that the $H_{\rm
FOT}$ and $H_{\rm SP}$ transition fields do persist down to a
system size of the order of hundred vortices. In addition, we
reveal that these features present the same signature and location
in the phase diagram, independently of decreasing the sample size
down to 21\,$\mu$m.

\section{Experimental}

Single-crystals of optimally-doped
Bi$_{2}$Sr$_{2}$CaCu$_{2}$O$_{8}$
 with $T_{\rm c}=90$\,K were grown by means of the traveling-solvent
floating zone technique \cite{Li94a}. We selected two high-quality
crystals from the same batch. The first one was taken as the
reference $220 \times 220 \times 30$\,$\mu$m$^2$ macroscopic
crystal and the second one was used to engineer micron-sized
disks. The latter are obtained by means of optical lithography of
the sample surface and subsequent physical ion-milling in the
disks' negative \cite{Dolz10a}. The resulting micron-sized towers
are removed by cleaving the milled surface with a resist-wetted
silicon substrate. The resist is then chemically eliminated. The
resulting circular samples have typical thicknesses of $1$\,$\mu$m
and diameters ranging from 10 to 50\,$\mu$m. The disks are then
mounted onto Hall-sensor chips with micron-precision manipulators,
and glued with Apiezon N grease.

\begin{figure}[ttt]
\includegraphics[width=\columnwidth,angle=0]{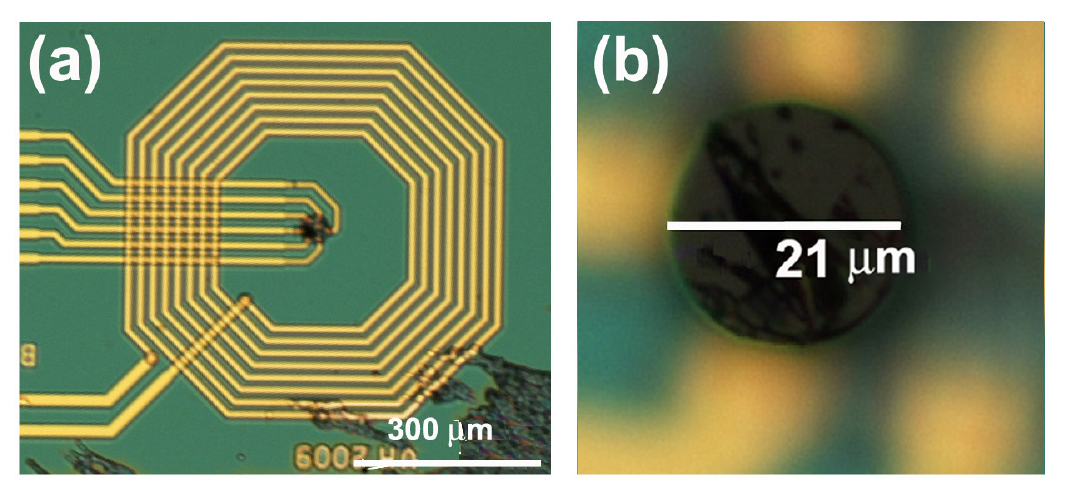}
\caption{Optical microscope image of a 21\,$\mu$m diameter
Bi$_{2}$Sr$_{2}$CaCu$_{2}$O$_{8}$ disk mounted on one of the
on-chip Hall sensors  embedded with the excitation coil. (a) Large
scale view of the two adjacent 2D electron gas Hall sensors, with
active areas of $6 \times 6$\,$\mu$m$^{2}$, and separated by
20\,$\mu$m. The coil  is used to generate a ripple AC field with
frequencies up to 1\,kHz. (b) Zoom of the sensor region with the
disk placed on the left sensor.  The surface of the sample is flat
such as this results from cleaving; the observed heterogeneities
on brightness are due to an irregular height of the Apiezon N
grease  used to glue the sample and to guarantee good thermal
contact with the sensors chip. \label{figure1}}
\end{figure}

\begin{figure}[ttt]
\includegraphics[width=\columnwidth,angle=0]{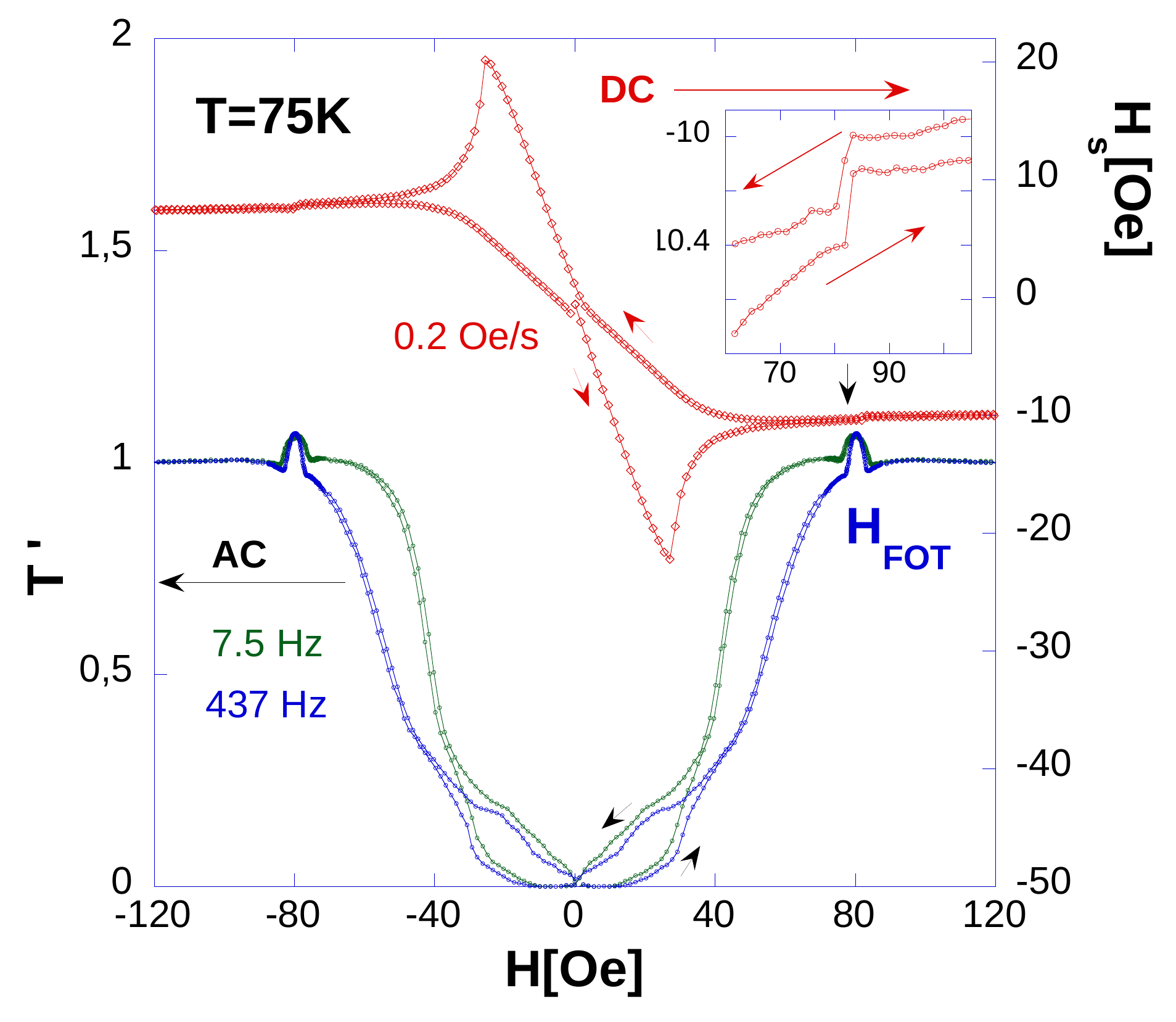}
\caption{High-temperature DC and AC magnetic hysteresis loops of
the reference Bi$_{2}$Sr$_{2}$CaCu$_{2}$O$_{8}$ macroscopic
crystal measured at 75\,K. Top: ascending and descending branches
of the DC hysteresis loop. Insert: zoom-in of the DC  loop at the
vicinity of the first-order transition entailing a $B$-jump (see
black arrow). Red arrows indicate the ascending and descending
branches. Bottom: AC transmittivity loops with paramagnetic peaks
fingerprinting the first-order transition field $H_{\rm FOT}$. The
loops were measured with ripple fields of 0.9\,Oe rms and
frequencies of 7.5 and 437\,Hz. Black arrows indicate the
ascending and descending branches. \label{figure2}}
\end{figure}

The local magnetization or, more precisely, the stray field, of
the disks was measured with microfabricated 2D electron gas Hall
sensors \cite{Konczykowski91a} with $6 \times 6$\,$\mu$m$^{2}$
active surface  embedded with an excitation coil on a single chip.
Figure \ref{figure1} shows a photograph of one of the studied
 disks mounted on one of two adjacent sensors. The remaining
empty sensor is used as a local reference of the applied magnetic
field when performing DC experiments. The macroscopic sample was
measured by means of  similar chips supporting several Hall
sensors
 with active areas ranging from $6 \times 6$ to $40
\times 40$\,$\mu$m$^{2}$. The Hall array is always placed over the
center of the sample. The on-chip embedded coil generates an AC
field parallel to the
 applied DC field $H$. In all the experiments presented here this
ripple field has an amplitude of $H_{\rm AC}=0.9$\,Oe rms, while
its frequency ranges from 1 to 1000\,Hz. The current applied to
the sensors is in the range of 25 to 50\,$\mu$A. A
digital-signal-processing  lock-in technique is used to
simultaneously measure the in- and out-of-phase components of the
fundamental and the third-harmonic frequencies of the Hall
voltage.

\begin{figure} [ttt]
\includegraphics[width=\columnwidth,angle=0]{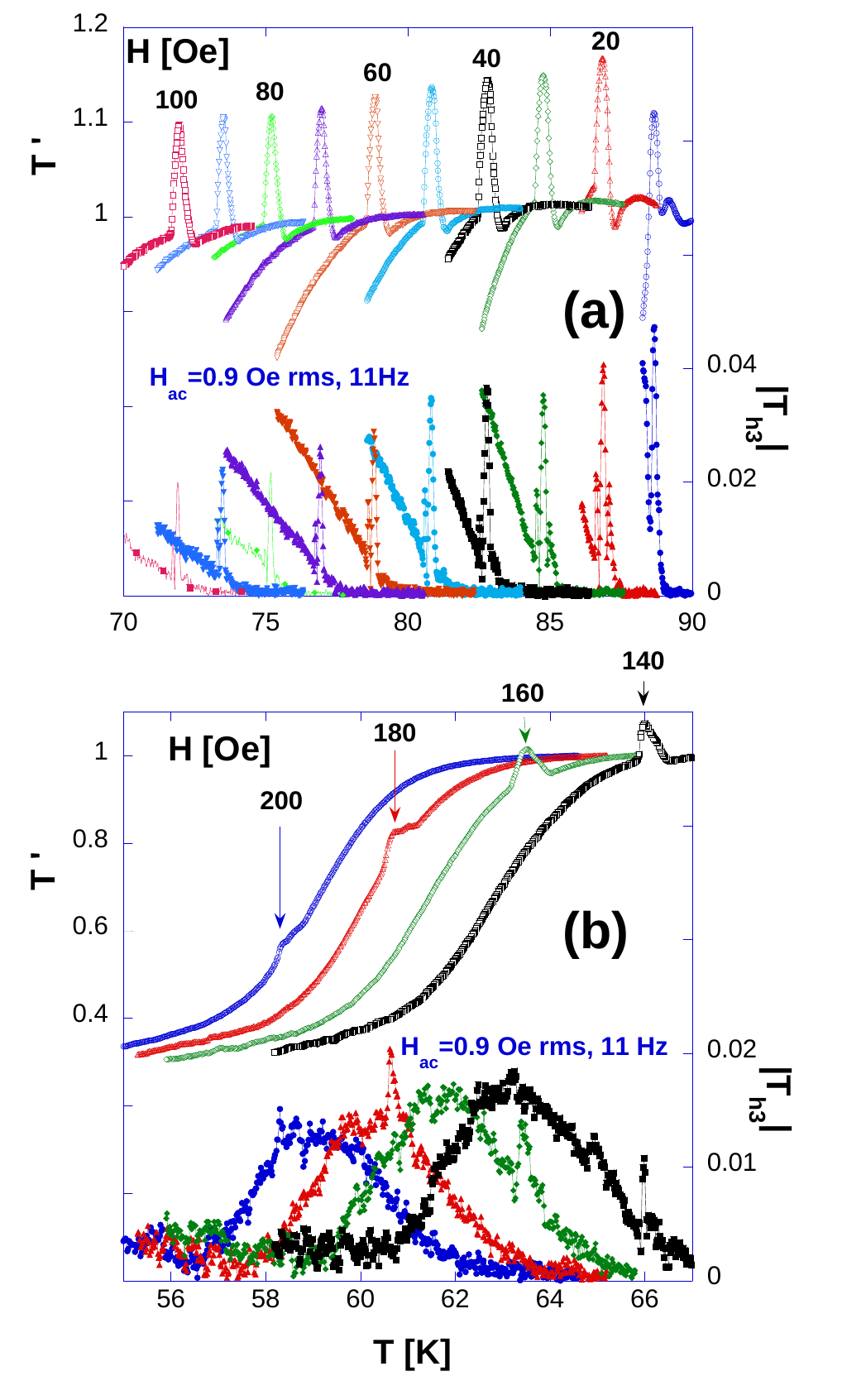}
\caption{Temperature-dependence of the transmittivity and modulus
of the third harmonic response for the reference
Bi$_{2}$Sr$_{2}$CaCu$_{2}$O$_{8}$ macroscopic crystal. (a) Low
field and (b) high field regimes. The AC ripple field of 0.9\,Oe
rms and 11\,Hz is collinear to the applied field $H$.
\label{figure3}}
\end{figure}

\begin{figure}
\includegraphics[width=0.9\columnwidth,angle=0]{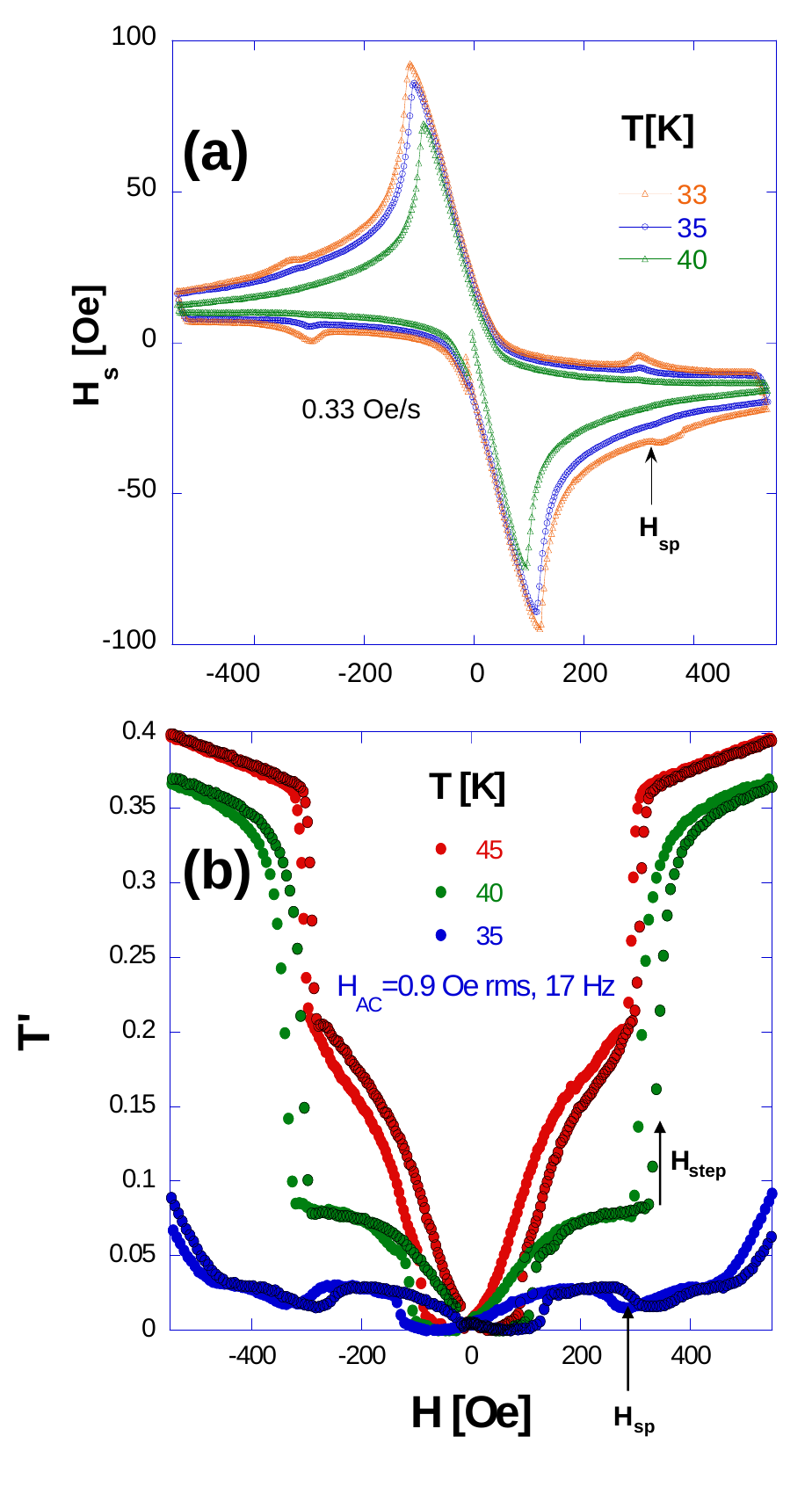}
\caption{DC and AC magnetic hysteresis loops in the
low-intermediate temperature regime for the
Bi$_{2}$Sr$_{2}$CaCu$_{2}$O$_{8}$ macroscopic crystal. (a) DC
loops from which the transition field $H_{\rm SP}$ is taken at the
midpoint between the onset and the full development of the local
minimum in the ascending branch (see arrow). (b) Transmittivity AC
loop measured with a ripple field of 0.9\,Oe rms and 17 Hz
parallel to $H$. At each temperature, black-on-color points
correspond to the positive ascending and negative descending field
branches of the transmittivity curve. The $H_{\rm SP}$  field is
obtained similarly as in DC loops (see arrow). The field location
of the step-like feature, $H_{\rm step}$, is taken at half the
step height in the ascending field branch (see arrow).
\label{figure5}}
\end{figure}

In this way, several DC and AC local magnetic measurements were
performed using the same set-up, as a function of temperature,
magnetic field, and frequency. The DC magnetic hysteresis loops
are obtained by measuring the magnetization, $H_{\rm S} = B - H$,
when cycling the static applied  field $H$. During the AC
measurements, an $H_{\rm AC}$ field collinear to $H$ is applied
and the first- and third-harmonic of the AC response of Hall
sensors is simultaneously recorded. The in-phase component of the
first-harmonic signal $B'$ is converted to the transmittivity $T'=
[B'(T) - B'(T \ll T_{\rm c})]/[B'(T>T_{\rm c}) - B'(T \ll T_{\rm
c})]$ \cite{Gilchrist1993}. This magnitude is more sensitive to
discontinuities in the local induction than direct measurements of
the static induction  $B$. The modulus of the third harmonic
signal $\mid T_{\rm h3} \mid= \mid B_{h3}^{\rm AC}
\mid/[B'(T>T_{\rm c}) - B'(T \ll T_{\rm c})]$ becomes non
negligible at the onset of irreversible magnetic properties (see
Fig.\,\ref{figure3}).  This onset is then used to track  $H_{\rm
irr}$ \cite{vanderBeek1995}.

We perform two types of measurements, temperature evolution of
$T'$ and $\mid T_{\rm h3} \mid$ on field-cooling in several
fields, and isothermal DC and AC hysteresis loops
\cite{Konczykowski2006} as, for instance, shown in
Fig.\,\ref{figure2}. By means of the AC technique, the first-order
transition field and the onset of magnetic irreversibility are
detected with better resolution than by DC loop measurements. In
the high-temperature regime, the FOT is manifest in the AC
transmittivity as a prominent paramagnetic peak that develops at
the same $H$ as the jump in local induction detected in DC
hysteresis loops\cite{Morozov1996,Konczykowski2006}.

\section{Results}

\subsection{Reference macroscopic sample}

The magnetic properties of the reference macroscopic
Bi$_{2}$Sr$_{2}$CaCu$_{2}$O$_{8}$ sample present three
characteristic temperature regimes. Figure\,\ref{figure2} depicts
the magnetic hysteretic response of the reference macroscopic
sample at high-temperature, $T = 0.83\, T_{\rm c}$. The top panel
shows a  two-quadrant DC loop such as typically obtained in this
temperature regime.  Closer inspection of the DC data in the
vicinity of 80\,Oe reveals a $H_{\rm S}$ jump with similar height
for both, ascending and descending branches. This feature is the
fingerprint of the first-order transition $H_{\rm FOT}$ in the
high-temperature region. The bottom panel of Fig.\,\ref{figure2}
shows that, in AC loops, this transition is detected with improved
resolution: paramagnetic peaks emerge at the flanks of the central
$T'$ depletion \cite{Konczykowski2006}. On increasing frequency,
the system enhances its shielding capability manifested as a $T'$
decrease. The field-location of the paramagnetic peak in AC loops
is frequency-independent, and is therefore considered as $H_{\rm
FOT}$.

Figure \ref{figure3} shows a set of AC magnetic data for low
applied fields ranging from 10 to 200\,Oe. The paramagnetic peak
in $T '$ shifts towards lower temperatures on increasing field.
The peaks are  sharp and  all transmittivity curves have the same
high-temperature background  up to 100\,Oe. On cooling at higher
fields, shielding currents develop in the sample prior to the
appearance of the paramagnetic peak. Nevertheless,
Fig.\,\ref{figure3} (b) shows that the paramagnetic peaks can be
detected up to 200\,Oe. The irreversibility line, identified from
the onset of the third-harmonic signal  on cooling, is located at
temperatures higher than the paramagnetic peak. This indicates the
existence of a phase region with non-linear vortex dynamics at
fields exceeding $H_{\rm FOT}$ \cite{Indenbom1996}. The amplitude
of $\mid T_{\rm h3} \mid$ decreases on increasing the applied
field, see Fig.\,\ref{figure3} (b).  In addition, the $\mid T_{\rm
h3} \mid$ curve echoes the paramagnetic peak in $T '$.

\begin{figure}
\includegraphics[width=\columnwidth,angle=0]{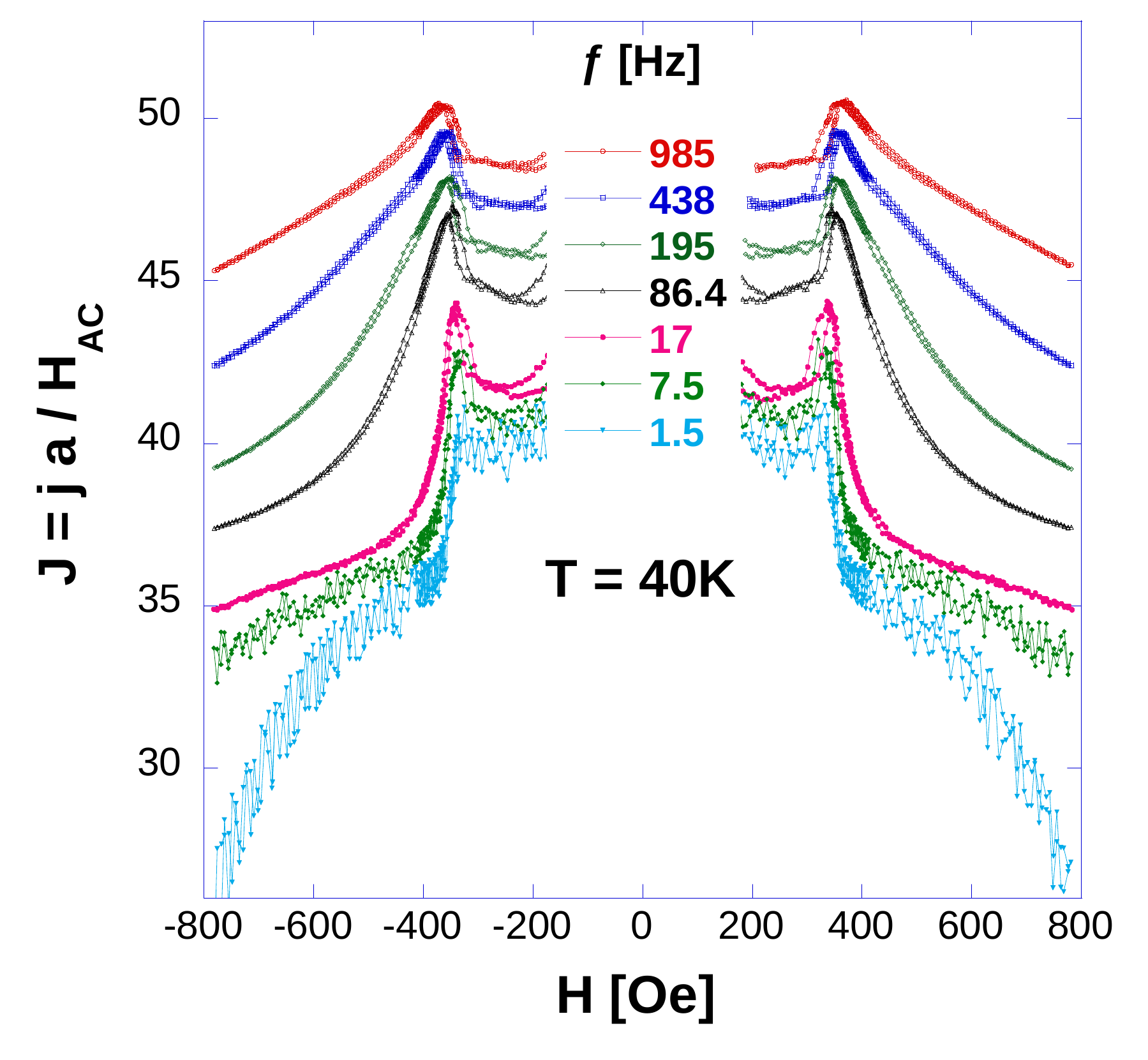}
\caption{Normalized sustainable-current density $J$ as a function
of applied field at 40\,K for the
Bi$_{2}$Sr$_{2}$CaCu$_{2}$O$_{8}$ macroscopic crystal. The
frequency dependence of $J$ is obtained from AC magnetization
loops measured with a ripple field of 0.9\,Oe rms and frequencies
ranging 1.5 to 985\,Hz. \label{figure6}}
\end{figure}

The  $H_{\rm SP}$ transition field can be obtained from  DC
magnetic hysteresis loops as those shown in Fig.\,\ref{figure5}
(a) for low-intermediate temperatures. In these curves the local
minimum (maximum) of the ascending (descending) branch becomes
more evident on cooling. The loops occupy two field-quadrants and
noticeably increase their width on cooling. This figure also
indicates that the $H_{\rm SP}$ field is hard to detected from DC
magnetic loops for temperatures higher than 33\,K. This is due to
a technical limitation of the DC technique that lacks resolution
in order to measure bulk transitions in the high-temperature
regime where surface barriers dominate the magnetic properties
\cite{Chikumoto1992}.

We therefore use the AC hysteresis loop technique in order to
track the first-order transition up to higher temperatures. The AC
transmittivity reflects the dimensionless normalized
sustainable-current density, $J=j(f)a/H_{\rm AC}$, provided $J =
(1/\pi)\arccos(2 T' - 1)$ \cite{vanderBeek1995}. This formula was
derived for an AC penetration in the Bean critical regime, an
assumption that seems to be valid in view of the results presented
in Ref.\,\onlinecite{Indenbom1996}. A low-temperature AC loop at
35\,K is shown in Fig.\,\ref{figure5} (b) depicting local minima
in both, the ascending and descending branches. These minima can
be ascribed to the second-peak transition $H_{\rm SP}$ since a
minimum in $T '$ corresponds to a maximum in the bulk $J$. The
$T'$ signal evolves in a different manner for the high- and
intermediate-temperature regimes. For temperatures larger than
$0.66\,T_{\rm c} =58$\,K, the transmittivity presents paramagnetic
peaks, developing at the flanks of the central depletion (see
Fig.\,\ref{figure2}). For intermediate temperatures a sudden jump
of $T'$ is detected, as for instance in the AC loops measured at
40 and 45\,K. This step-like feature, manifested at a field
$H_{\rm step}$ implies a frequency-independent  drop of the
magnetic hysteresis, consistent with the change in the width of
the DC loops, and indicating that this transition is governed by
the local value of $B$ rather than $H$ \cite{Khaykovich1996}.  The
jump in $T'$ is related to the sudden change in shielding currents
with the sample becoming more transparent to the penetration of
the AC ripple field at larger $H$. Similarly, the height of the
steps in $T '$ decreases on increasing temperature.

The crossover temperature for the detection of the second-peak or
of the step-like feature is time/frequency dependent.
Figure\,\ref{figure6} presents the evolution of the normalized
sustainable-current density extracted from $T'$ as a function of
frequency in the range from 1.5 to 985\,Hz. For frequencies
smaller than 7.5\,Hz a sudden drop of $J$ develops at fields $
H_{\rm SP} \sim 330$\,Oe. On increasing frequency, this feature
evolves into a field-asymmetric peak producing the step-like
feature observed in $T'$ curves. The onset of this peak, or
equivalently the $H_{\rm step}$ field in $T'$, is frequency
independent.

\begin{figure}
\includegraphics[width=\columnwidth,angle=0]{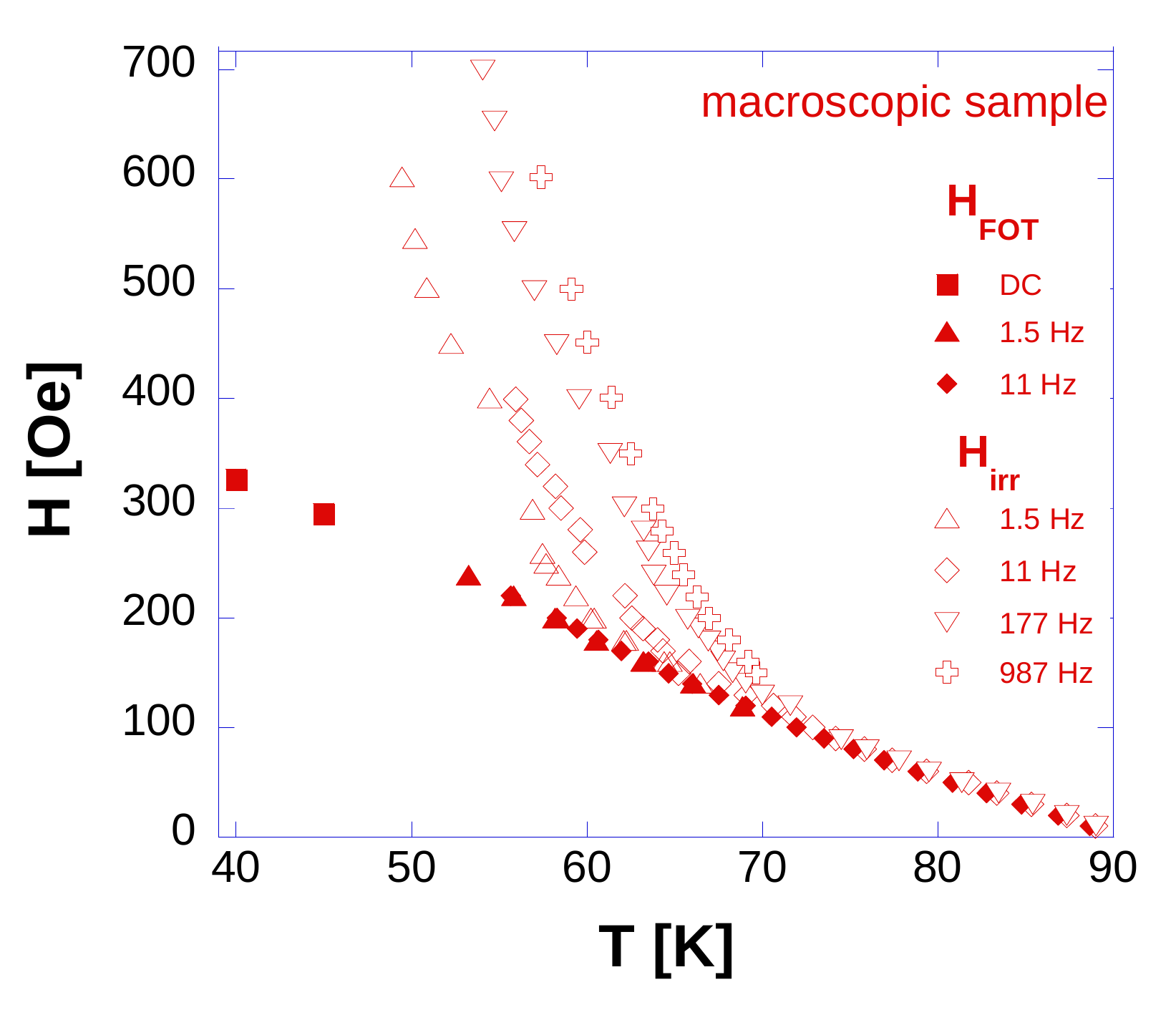}
\caption{Vortex phase diagram of the macroscopic reference
Bi$_{2}$Sr$_{2}$CaCu$_{2}$O$_{8}$ sample. First-order transition
(full points) and irreversibility (open points) lines from DC and
AC magnetic measurements at different frequencies. At low
temperatures, the $H_{\rm irr}$ line strongly depends on
frequency. Measurements were performed at an AC ripple field of
0.9\,Oe rms. \label{figure7}}
\end{figure}

Finally, Fig.\,\ref{figure7}  presents the vortex matter phase
diagram of the macroscopic Bi$_{2}$Sr$_{2}$CaCu$_{2}$O$_{8}$
sample in the high-temperature region, indicating the $H_{\rm
FOT}$ (full points) and the $H_{\rm irr}$ (open points) lines.
Both lines merge for $T>70\,K=0.77\,T_{\rm c}$, whereas at smaller
temperatures, the $H_{\rm irr}$ line deviates towards higher
fields and is strongly frequency-dependent. The higher the
frequency, the larger is the deviation from the FOT line. The
onset of $\mid T_{\rm h3} \mid$ indicates the electric field $E =
E(H_{\rm AC},f, J)$ becomes non-linear below a working point given
by the experimental resolution. As temperature decreases, the
linear response limit shrinks to  smaller current-density values.
Therefore, the onset of third harmonic response shifts to lower
temperatures for smaller frequencies (smaller currents).

\subsection{Micron-sized samples}

\begin{figure}[ttt]
\includegraphics[width=\columnwidth,angle=0]{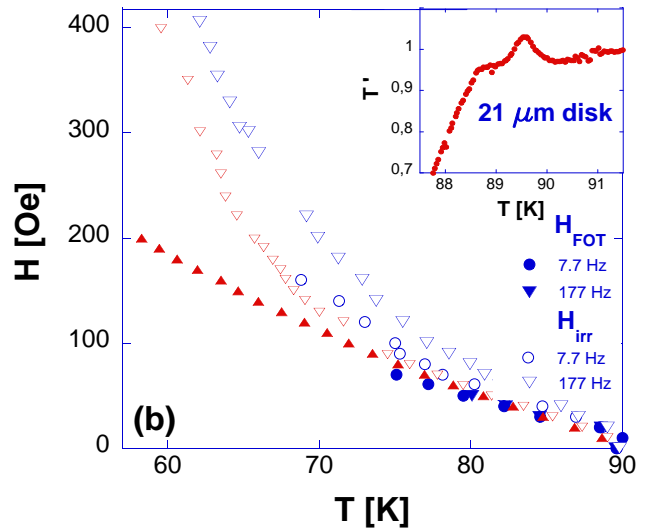}
\caption{Vortex phase diagram of  micron-sized
Bi$_{2}$Sr$_{2}$CaCu$_{2}$O$_{8}$ vortex matter for the smallest
measured disk of 21\,$\mu$m diameter (blue points). First-order
transition (full points) and irreversibility (open points) lines.
Data from the macroscopic reference sample (red points) are
included for comparison. Inset: Low-field transmittivity for the
smallest measured field of 5\.Oe. Measurements were performed at
an AC ripple field of 0.9\,Oe rms and 7.7 and 177\,Hz.
\label{figure11}}
\end{figure}

As discussed in the previous section,  performing AC  measurements
is necessary in order to properly track the FOT line, particularly
in the case of samples of highly-reduced size in which surface
barriers for vortex flux entry/exit completely dominate the
electromagnetic response. The insert to Fig.\,\ref{figure11} shows
the transmittivity data for the 21\,$\mu$m diameter disk at the
smallest measurement field of 5\,Oe. The paramagnetic peak
fingerprinting the $H_{\rm FOT}$ is clearly visible for this
system consisting of only 80 vortices. Tracking the location in
the temperature-field plane of this peak yields the first-order
transition line in the vortex phase diagram shown in the main
panel of Fig.\,\ref{figure11}. An important finding of this figure
is the persistence of the $H_{\rm FOT}$ transition line for
micron-sized vortex matter from $T_{\rm c}$ down to
$75\,K=0.83\,T_{\rm c}$ . The first-order transition line is also
detected down to similar temperatures in larger disks. This
indicates that, even for samples with  a large surface-to-volume
ratio, the $H_{\rm FOT}$ transition measured at high-temperatures
remains robust. For $T < 75\,$K no paramagnetic peak is detected
in the transmittivity data, presumably due to a masking effect
produced by  enhanced surface barriers in small samples. The
$H_{\rm FOT}$ line for the disks merges that of the macroscopic
reference sample, indicating that the nature of this transition
does not change even for a system with just 80 vortices, for the
smallest applied field of 5\,Oe in the 21\,$\mu$m sample (see
Fig.\,\ref{figure11}).

The $H_{irr}$ line for micron-sized samples is located at higher
fields than for the macroscopic sample. For example,
Fig.\,\ref{figure11} shows that at low temperatures $H_{\rm irr}$
is $\sim 35\%$ larger for the 21\,$\mu$m disk than for the
macroscopic sample at a fixed frequency (177\,Hz in this case).
This result might have origin in the different aspect ratio of the
macroscopic and disk samples, and in their probably dissimilar
surface roughness originating from the different preparation
methods for both specimens. Ascertaining the origin of this
discrepancy is beyond the aim of this paper. Figure \ref{figure11}
also shows that for smaller frequencies $H_{\rm irr}$ approaches
the first-order transition line. Finally, in the high-temperature
range the irreversibility lines for the macroscopic and
micron-sized samples merge into a single bunch of data with the
$H_{\rm FOT}$ line.

\begin{figure}
\includegraphics[width=\columnwidth,angle=0]{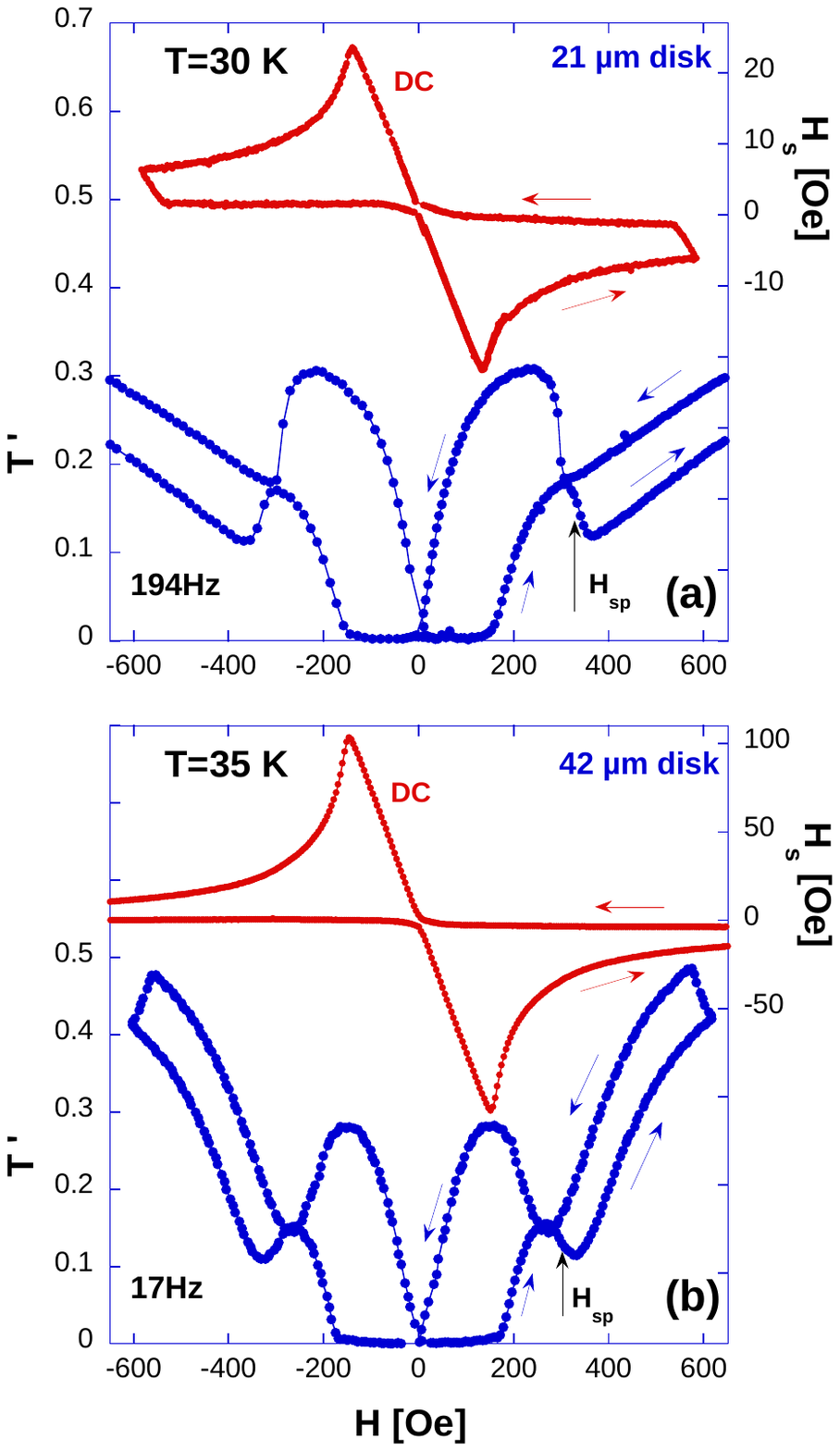}
\caption{Low-temperature magnetic hysteresis loops for
Bi$_{2}$Sr$_{2}$CaCu$_{2}$O$_{8}$ micron-sized disks for DC (top
red curves) and AC (bottom blue curves) measurements: (a) smallest
21\,$\mu$m diameter measured disk at 30\,K and (b)  42\,$\mu$m
diameter disk at 35\,K. The ripple field is of 0.9\,Oe rms and
194\,Hz for the smaller disk and 17\,Hz for the largest one. The
transition field $H_{\rm SP}$ considered as the mid-point between
the onset and the full development of the minimum in the ascending
branch is indicated with an arrow.   \label{figure8}}
\end{figure}

We also study the effect of decreasing the sample size down to
microns on the $H_{\rm SP}$ transition. Previous works using DC
magnetic techniques reported that this transition field is not
detected for samples with diameters smaller than 30\,$\mu$m
\cite{Wang2000,Wang2001}. We performed AC measurements  in order
to study the persistence or absence of the $H_{\rm SP}$
transition. Fig.\,\ref{figure8} shows the comparison of DC and AC
loops data: while the DC loops do not show evidence of the $H_{\rm
SP}$ feature, the local minima in the ascending and descending
branches of the AC loops are evident. Even measuring at
intermediate (194\,Hz) and low (17\,Hz) frequencies, the
application of the AC technique enables the observation of the
emergence of the bulk current contribution, therefore allowing the
detection  of $H_{\rm SP}$. Figure \ref{figure8} depicts a
remarkably well-developed $T '$ local minimum for the 21 and
42\,$\mu$m diameter disks, giving a $H_{\rm SP}$ field that is
within 10\% of the value observed in macroscopic samples (roughly
330\,Oe).

Therefore, the results of Fig.\,\ref{figure8} show that the
previously reported absence of the $H_{\rm SP}$ transition for
micron-sized disks of Bi$_{2}$Sr$_{2}$CaCu$_{2}$O$_{8}$ is  the
consequence of a technical limitation. The mentioned
works\cite{Wang2000,Wang2001} apply DC magnetic techniques
entailing large characteristic measurement times. In this case,
surface currents play a determining role in screening, as already
known from studies on macroscopic
Bi$_{2}$Sr$_{2}$CaCu$_{2}$O$_{8}$ crystals \cite{Chikumoto1992}.

\section{Discussion}

The three characteristic temperature regimes of the first-order
phase transition of \textit{macroscopic}
Bi$_{2}$Sr$_{2}$CaCu$_{2}$O$_{8}$  vortex matter are also found
for \textit{micron-sized} samples.  In the latter case,  the
identification of the first-order transition is rendered more
difficult by the enhancement of the surface to bulk-currents
ratio. The nearly-vanishing remanent magnetization and the
two-quadrant locus of the DC loops in \textit{micron-sized}
samples indicate the preeminence of surface barriers for vortex
flux entry/exit. Therefore applying AC magnetic techniques is
imperative in order to have access to the faster decaying bulk
currents emerging from the surface-barrier background
\cite{Chikumoto1992}.

In the intermediate- and low-temperature regimes  $T'$ and $\mid
T_{\rm h3} \mid$ present  particular features that are related to
the bulk-current contribution. For $0.39\,T_{\rm c} < T <
0.66\,T_{\rm c}$ we detect, at a field $H_{\rm step}$, a
discontinuous decrease of the bulk shielding-currents associated
with the first-order transition . On
 cooling below $T=0.39\,T_{\rm c}=35$\,K,
 the opposite effect of an increase of the
shielding currents is observed at almost the same field indicating
the $H_{\rm SP}$ transition. In the vicinity of
 this reversal of current behavior, varying the frequency of the AC
 ripple field tunes  a decrease (low-frequencies) or an
 increase (high-frequencies) of the shielding currents. The latter
 case is equivalent to probing magnetic relaxation on a shorter
 time-scale,
 in analogy to DC magnetization experiments \cite{Chikumoto1992}, or to
 choosing
a higher electric field in an transport $I(V)$ measurement. Since
the $I(V)$ curves just below and above the FOT cross --- with the
electric field in the high-field phase being larger than that in
the low-field phase for the low-current density limit, and
vice-versa for the high-current density limit ---, varying the
working point (by tuning the frequency) leads to either a step
like-behavior of the screening current (at low electric fields) or
a peak-like curve (at high electric fields)
\cite{Konczykowski2000}.
 The energy barriers for flux creep have
$U(J)$ and $E(J)$ curves with different functionalities for fields
larger or smaller than the transition one. On varying field close
to the transition these curves cross, and the  phase transition
produces a discontinuous change on the electrodynamics of vortex
matter. Detecting the transition with a high electric field (short
measurement times) leads to an enhancement of shielding currents
with increasing field, whereas with a low electric field (long
times) a sudden decrease is observed. The lack of sensitivity of
DC magnetization loops to the FOT in the intermediate-temperature
regime can be understood as the consequence of a distribution of
Bean-Livingston surface barriers \cite{Morozov1996b}. The transit
of  individual pancake vortices over these barriers is not
affected by the nature of the vortex phase inside the sample.

The irreversibility line obtained from the onset of $\mid T_{\rm
h3} \mid$ merges with the first-order transition line $H_{\rm
FOT}$ in the high-temperature regime. At lower temperatures or
higher applied fields, the shielding of the AC field starts at
higher temperatures than the occurrence of the first-order
transition, namely, the onset of $\mid T_{\rm h3} \mid$ develops
before the paramagnetic peak on cooling. This indicates the
existence of a non-linear vortex region spanning at higher
temperatures than the first-order transition line
\cite{Indenbom1996}. This phenomenon might have origin from a
residual effect of pinning \cite{Vinokur1991}, or Bean-Livingston
barriers \cite{Fuchs1998}on the high-temperature liquid phase.
This phenomenology is observed in \textit{macroscopic} as well as
 \textit{micron-sized} Bi$_{2}$Sr$_{2}$CaCu$_{2}$O$_{8}$
samples.

The most important finding of this work is that the first-order
transition persists even for a very small vortex system. In the
high-temperature regime, the clear detection of the paramagnetic
peak at 5\,Oe in the 21\,$\mu$m disk is shown in the insert to
Fig.\,\ref{figure11}. This result indicates that, independently of
its nature, the first-order transition is robust and its
thermodynamic nature remains unaltered even for a reduced system
size of less than hundred vortices. In the low-temperature regime,
the detection of the $H_{\rm SP}$ transition as a sudden increase
of shielding currents (decrease of $T'$) persists in
\textit{micron-sized} samples. The observation of this feature was
made possible by the use of an AC magnetic measurement technique.
The possibility of working at shorter time-scales (larger
frequencies) reveals  currents flowing in the sample volume with
improved sensitivity than DC measurements. Therefore, our results
indicate that reducing the sample size down to few dozens of
microns does not produce a disappearance nor a dramatic decrease
of $H_{\rm SP}$ due to size effects as previously claimed
\cite{Wang2000,Wang2001,Kalisky2005}.

\section{Conclusions}

In the high-temperature phase region, the first-order transition
of Bi$_{2}$Sr$_{2}$CaCu$_{2}$O$_{8}$ vortex matter remains robust
and persists at the same $H-T$ location  even when decreasing the
system size down to less than hundred vortices. We found that the
second-peak transition detected in the low- and
intermediate-temperature regions  also persists on decreasing the
system size down to roughly 20\,$\mu$m, in contrast with previous
reports \cite{Wang2000,Wang2001}. The identification of these
transitions increases in difficulty when decreasing the sample
size, due to the predominance of surface-barrier-related currents.
The application of AC magnetic techniques, with better sensitivity
to
 bulk currents, allowed us to
detect the transition-related features.

\section{Acknowlodgements}

We thank C.J. van der Beek for selecting the crystals and
discussing the manuscript. This work was supported by the ECOS
Sud-MINCyT France-Argentina collaboration program, Grant No.
A09E03. Work done at Bariloche was partially funded by PICT-PRH
2008-294 from the ANPCyT.

$^*$ To whom correspondence should be addresed:
yanina.fasano@cab.cnea.gov.ar

\end{document}